\documentclass[twocolumn]{svjour3}         
\smartqed  
\usepackage{graphicx}
\usepackage{mathptmx}      
\journalname{Astrophysics and Space Science}
\begin{document}

\title{The H$_{2}$ density within spiral and irregular galaxies 
at high redshift: estimating CO detection limits}

\titlerunning{The H_{2} density within galaxies at high redshift}        

\author{Mercedes Moll\'{a} \and Eduardo Hardy  \and 
\'{A}ngeles I. D\'{\i}az    
}

\authorrunning{Moll\'{a}, Hardy \& D\'{\i}az} 

\institute{M.Moll\'{a} \at
  Astrof\'{\i}sica de Part\'{\i}culas, CIEMAT, Avda. Complutense 22, \\ 
  28040-Madrid (Spain),  \email{mercedes.molla@ciemat.es}           
\and
  E. Hardy \at
  NRAO,  Casilla El Golf 16-10, Santiago (Chile), \\
  \email{ehardy@nrao.edu} \\
  \footnotesize{\sl The National Radio Astronomy Observatory is a 
   facility of the National Science Foundation operated under cooperative 
   agreement by Associated Universities, Inc. } 
\and
   A. I. D\'{\i}az \at
   Dpto. de F\'{\i}sica Te\'{o}rica, Universidad Aut\'{o}noma de Madrid,\\ 
   Cantoblanco, 28049-Madrid (Spain),  \email{angeles.diaz@uam.es}   
}

\date{Received: date / Accepted: date}

\maketitle

\begin{abstract}
We have computed a grid of chemical evolution models for a large set
of spiral and irregular theoretical galaxies of different total
mass. In our models, the gas phase has two components, the diffuse and
the molecular one ($\rm H_{2}$). It is possible, therefore, to follow
the time (or redshift) evolution of the expected density of the $\rm
H_{2}$ phase. We will show the predictions of this gas density at
hight redshift, which might be detected with ALMA, in this type of
galaxies.

\end{abstract}

\section{Introduction}
\label{intro}

The molecular interstellar medium plays a critical role in the
evolution of galaxies since it provides the material from which stars
form. There are observations \cite{sol05} of rotational transitions of
CO within $1.0 < z <6.5$, mostly in quasars and Early Universe
Molecular Emission Line Galaxies (hereinafter QSO's and EMGs,
respectively) which demonstrate that molecular clouds exist already at
the epoch of the galaxy formation. However, the current sample is
scarce for $z > 3 $, where the study of the star and galaxy formation
is specially interesting. ALMA is the instrument that may solve this
problem.

In our chemical evolution models the gas phase has two components, the
diffuse and the molecular phases of gas, which we treat
separately. This way, it is possible to follow the evolution of the
expected density of H$_{2}$ with redshift.  We will show here our
predictions about the $\rm H_{2}$ density at high redshift, and we
will analyze if they might be detectable with ALMA.

\section{A grid of multiphase chemical evolution models }
\label{sec:models}

We have computed a wide grid of chemical evolution models for a large
set of theoretical spiral and irregular galaxies of different total
mass \cite{mol05} by using the multiphase chemical evolution model
from \cite{fer92,fer94,mol96}.  We have simulated theoretical models
with 44 different radial distributions of total mass and 10
star-forming efficiencies, with values in the range [1--0], which we
denoted by $N$ from 1 to 10, for each one of them. The 440 resulting
models provide the time evolution for specific radial regions within
each galaxy modeled, with the corresponding mass in each gaseous or
stellar phase, as well as their elemental abundances. This type of
model was already applied to several individual spiral galaxies in
\cite{mol97}, where it was shown that the model is adequately
calibrated and that it may be used for galaxies of different
morphological types and different total masses.

The code assumes that a total mass, $M_{gal}$, initially as diffuse
gas, is located within a protogalaxy or halo. The gas falls onto the
equatorial plane forming out the spiral disk at a rate defined by a
collapse time-scale which depends on $M_{gal}$. In each zone, halo or
disk, several radial regions are defined. We follow separately the
evolution in each region.

The assumed star formation in the halo is a power law of the diffuse
gas.  In the disk, however, we assume that the star formation occurs
in two steps: first, molecular clouds ($\rm H_{2}$) form from the
diffuse gas (HI) also following a power law.  
\begin{figure*}
\centering
\includegraphics[width=0.65\textwidth, angle=-90]{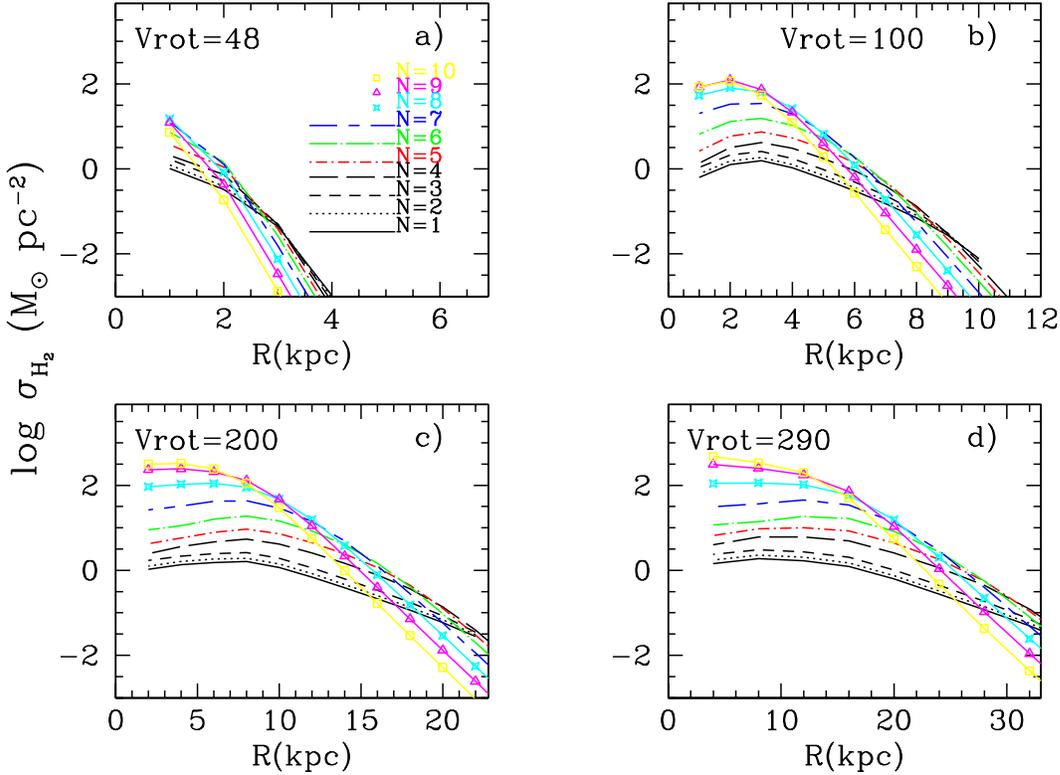}
\caption{Radial distributions of molecular gas surface density at $z =
0$. Each panel represents the results for a different rotation
velocity galaxy, which means a different galaxy total mass. Curves are
for different star formation efficiencies in the range 1 (the highest
efficiency, $N=1$) to 0 (the lowest efficiency, $N=10$) as labelled in
panel a).}
\label{fig1}
\end{figure*}Then, stars form by two
means: 1) cloud-cloud collisions, and 2) by the interaction of the
massive stars with the surrounding molecular clouds: This way, in our
models the gas has two components: HI and $\rm H_{2}$.

Since we have computed separately the diffuse gas evolution and the
molecular clouds evolution, we obtain the radial distributions of the
molecular gas mass, which is represented in Fig.~\ref{fig1}. There we
show the radial distributions of the molecular gas density at the
present-time for 4 values of the rotation velocity, or total
mass of the Galaxy. Thus, panel a) shows the radial distributions
produced in a low mass galaxy while d) represents a massive
galaxy. Panel c) is the modeled galaxy most similar to our Milky Way
Galaxy and panel d) is an intermediate mass galaxy. In each one
of these panels models for different star formation efficiencies,
named by $N$, are represented.  The lines in color correspond to the
lowest star-forming efficiencies (high $N$ values), which maintain a
higher density than in cases where stars are formed more efficiently
(black lines). The distributions are close to exponential in the
external regions but are flatter at the inner regions, mostly when the
N is low, as is in fact observed in Nature \cite{nish01}.

\begin{figure*}
\includegraphics[width=0.65\textwidth,angle=-90]{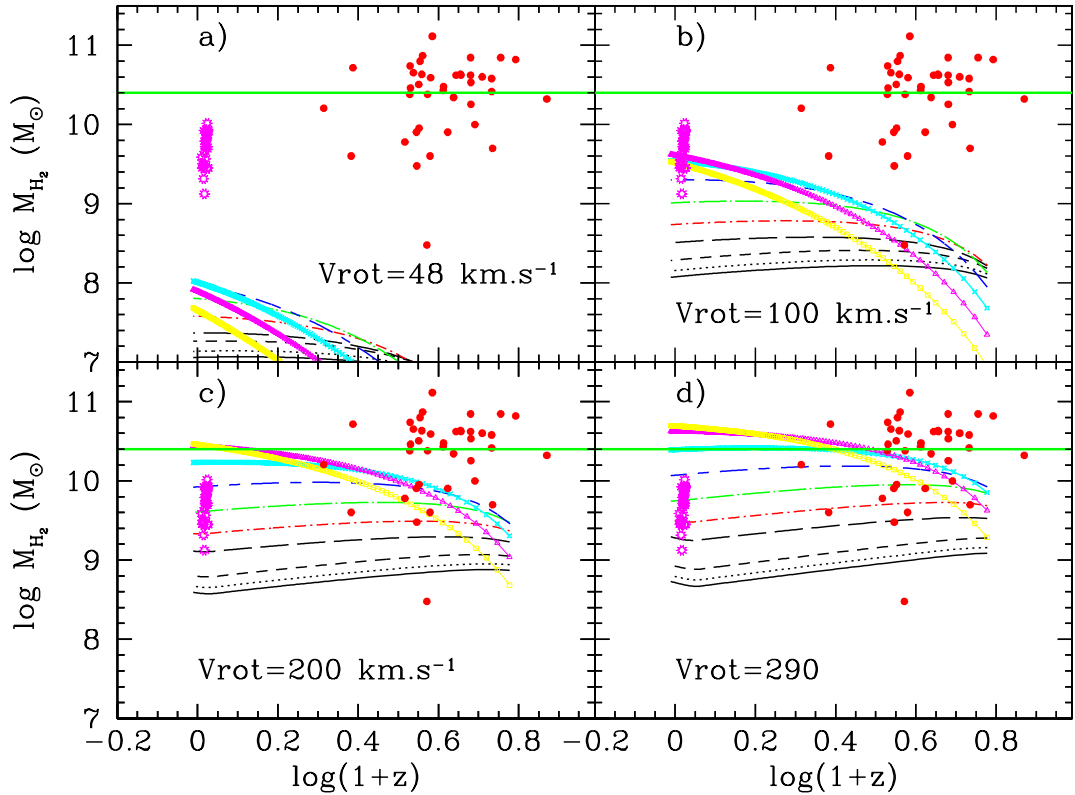}
\caption{ The evolution of the total mass of molecular gas for some of
our models. Each panel represents the results for a given galaxy total
mass (or Vrot), using, for each one, 10 different SFR
efficiencies. Lines mean the same as in Fig.1 The black lines
correspond to the highest efficiencies.}
\label{fig2}
\end{figure*}

The evolution of the total mass of $\rm H_{2}$ in galaxies is
represented in Fig~\ref{fig2}. The magenta open stars are estimates from
\cite{ber05} for nearby QSOs. The predicted present time molecular gas
masses are similar to those ones of local objects with a maximum for
$z=0$ in agreement with the observational limit $\rm
log(MH_{2})=10.4$, shown by a green solid line, obtained for those
authors. The red full dots are the estimates from \cite{sol05} for
EMGs. The observed molecular gas masses are decreasing for decreasing
$z$ in agreement with the most of our model results.  Only the models
with low star-forming efficiencies (colored lines) are still producing
$\rm H_{2}$.

On the other hand, following that work \cite{sol05}, and taking into
account the estimates for the star formation rate and molecular gas
total mass, these objects might be bulges or starbursts in centrally
concentrated disks, but not elliptical galaxies as it was initially
assumed.  Our models are also in agreement with this statement: the
total mass of $\rm H_{2}$ for the most massive spiral galaxies (with
$Vrot$ in the range 100-300 km.$\rm s^{-1}$) and high star formation
efficiencies ($N < 5$) are always below of the EMGs data region since
in those cases the molecular gas is very quickly consumed.  Only the
models for massive galaxies -panels c) and d) with the lowest
efficiencies show values closer to the observations of EMG's. At this
high redshift, these particular models show star formation only at the
central regions and therefore they would appear as concentrated disks,
with a large quantity of $\rm H_{2}$ .

\section{The CO detection limits}

Observation of emission from CO rotational transitions is the usual
means of detecting interstellar molecular clouds. 
\begin{figure}
\centering
\includegraphics[width=0.45\textwidth,height=0.3\textheight,angle=0]{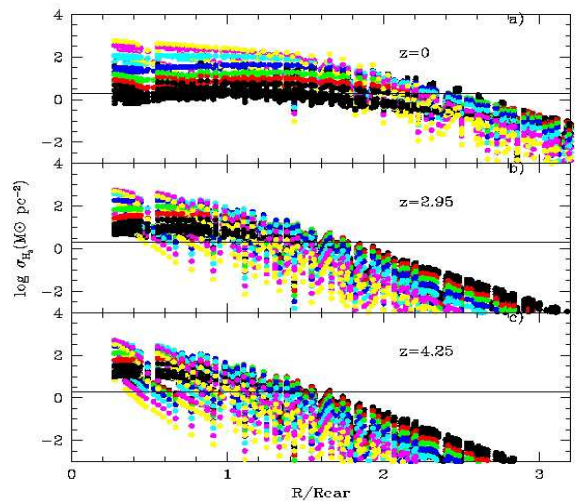}
\caption{The expected radial distributions of molecular gas density
normalized to $Rcar$ (half optical radius) for three 
redshifts as labelled.}
\label{fig3}
\end{figure}
CO is a very stable molecule and the most abundant after
H$_{2}$. Following \cite{com01} levels $J=4-5$ are the best ones to
detect molecular gas at high redshifts.

In order to verify if our densities of molecular gas may be detected
with ALMA, we need to transform the limiting flux for this instrument
to a limit in gas density. The flux limit depends on the observed
frequency. We take the highest value from the report "Science with
ALMA", and we use Eq. 1 from \cite{sol05}, to calculate the minimum
L$_{\rm CO}$ necessary for detection. This luminosity is transformed
into gas mass by using, as those authors, a constant conversion factor
$\alpha=0.8$ (which assumes independence of redshift $z$).
\begin{figure*}
\centering
\includegraphics[width=0.65\textwidth,angle=-90]{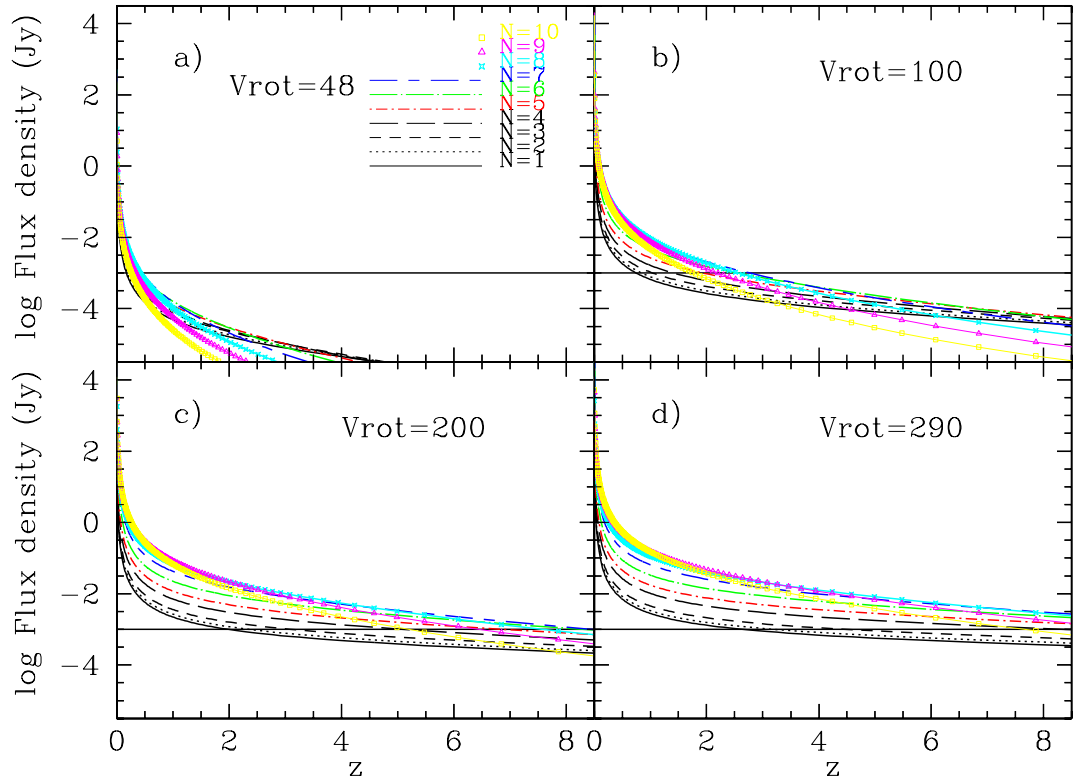}
\caption{Expected evolution of the flux density of CO lines as a
function of z for spiral galaxies with four rotation velocities. Line
meaning as in Fig.1}
\label{fig4}
\end{figure*}
The result is shown as a thick line in Fig~\ref{fig3}, where we show
the expected radial distributions of molecular gas density normalized
to a characteristic radius $Rcar$, taken as the half optical radius,
for each total mass radial distribution. We show in top panel a) the
results for $z=0$ (present time) while panels b) and c) represents the
same models for the higher redshifts $z=2.95$ ($t\sim 1.2$ Gyr) and
$z=4.25$ ($t\sim 0.6$ Gyr), respectively.  For $z=0$ most modeled
galaxies show distributions with values higher than the limiting value
which may be detected with ALMA down to the optical radius, while for
high redshifts only our inner disks, mainly of low efficiencies (colored)
models, have a flux density sufficiently high to be
detected. Therefore, if the red points of Fig.~2 would correspond to
spiral galaxies, they must represent, effectively, their central
parts.

In Fig.~\ref{fig4} we represent the expected evolution of the flux
density of CO lines as a function of $z$, for an integrated galaxy,
computed from the molecular gas density $\rm \sigma (H_{2}$) obtained
from our models, transformed to fluxes by using the method outlined.
The most massive galaxies -panel d- and those with high star formation
efficiencies -black lines-- maintain an almost constant $\rm H_{2}$
density and flux density up to $ z \sim 10$, with values just over the
detection limits. Thus, such as we have before claimed, the most
efficient galaxies in forming stars, which consume very quickly the
molecular gas, are not the most probable objects to be detected. Those
with low rotation velocities $Vrot < 50$ km.s-1, on the other hand,
show increasing densities for decreasing z, having values lower than
the limiting flux for $z > 0.5$.  Only the massive galaxies $Vrot>
200$ km.s$^{-1}$ with $\rm N>5$ have fluxes above the
observational limit even at $z \sim 8$.  Therefore, the most plausible
objects to be observed with ALMA will be the centers of
low-intermediate efficiencies massive galaxies.

\section{Conclusions}
\begin{itemize}

\item At redshift $z > 3$ only the central regions
of galaxies might be observed in CO with ALMA

\item EMGs are not necessarily the largest or the most massive stellar disks.
\end{itemize}


\begin{thebibliography}{}

\bibitem{ber05}
Bertram, T., Eckart, A., Krips, M., Straubmeier, C., Fischer, S., \&
Staguhn, J.~G.\ 2006, New Astronomy Review, 50, 712

\bibitem{com01}Combes, F.\ 2001, SF2A-2001:
Semaine de l'Astrophysique Francaise, 237

\bibitem{fer92}
{Ferrini} F.,  {Matteucci} F.,  {Pardi} C.,    {Penco} U.,  1992, ApJ, 387, 138

\bibitem{fer94}
{Ferrini} F.,  {Moll\'{a}} M.,  {Pardi} M.~C.,    {D{\'{\i}}az} A.~I.,  1994,
  ApJ, 427, 745

\bibitem{mol05}
{Moll{\' a}} M.,  {D{\'{\i}}az} A.~I.,  2005, MNRAS, 358, 521

\bibitem{mol96} 
{Moll\'{a}, M., Ferrini, F., \& D\'{\i}az, A. I., 1996, ApJ, 466, 668}

\bibitem{mol97} 
{Moll\'{a}, M., Ferrini, F., \& D\'{\i}az, A. I., 1997, ApJ, 475, 519}

\bibitem{nish01}
{Nishiyama, K. ~\& Nakai, N., 2001, PASJ, 53, 713}

\bibitem{sol05}
Solomon, P.~M., \& Vanden Bout, P.~A.\ 2005, ARA\&A, 43, 677

\end{thebibliography}
\end{document}